\documentclass[a4paper,11pt]{article}
\pdfoutput=1
\usepackage{jheppub}
\usepackage[T1]{fontenc}
\usepackage{color}
\usepackage{amssymb,amsmath,comment}
\usepackage{graphicx}
\usepackage{braket}
\usepackage{epsfig}
\usepackage{here}
\graphicspath{{./Figures/}}
\newcommand{\ba}{\begin{align}}
\newcommand{\ea}{\end{align}}

\def\nn{\nonumber}

\def\LL{\mathrm{LL}}

\def\RR{\mathrm{RR}}

\def\bea{\begin{eqnarray}}
\def\eea{\end{eqnarray}}
\title{\boldmath Direct CP Violation in Cabibbo-Favored Charmed Meson Decays and $\epsilon'/\epsilon$\\
in $SU(2)_L\times SU(2)_R\times U(1)_{B-L}$ Model}
\author[]{Naoyuki Haba,}
\author[1]{Hiroyuki Umeeda\note{Corresponding author.}}
\author[]{and Toshifumi Yamada}

\affiliation[]{Institute of Science and Engineering, Shimane University, Matsue 690-8504, Japan}

\emailAdd{haba@riko.shimane-u.ac.jp}
\emailAdd{umeeda@riko.shimane-u.ac.jp}
\emailAdd{toshifumi@riko.shimane-u.ac.jp}

\abstract{Since the standard model contribution is virtually absent, 
 any observation of direct CP violation in the Cabibbo-favored charmed meson decays would be evidence of new physics.
In this paper, we conduct a quantitative study on direct CP violation in $D^0\to K^-\pi^+$, $D_s^+\to \eta\pi^+$ and $D_s^+\to \eta'\pi^+$ decays
 in the $SU(2)_L\times SU(2)_R\times U(1)_{B-L}$ gauge extension of the standard model.
In the model, direct CP violation arises mainly from the interference between the decay amplitude coming from the SM left-left current operators and 
 that from the right-right current operators induced by $W_R^+$ gauge boson exchange.
Interestingly, the strong phase between the two amplitudes is evaluable, 
 since it stems from difference in QCD corrections to the left-left 
 and right-right current operators, which is a short-distance QCD effect
 given by $\sim(\alpha_s(M_{W_L}^2)/4\pi)\log(M_{W_R}^2/M_{W_L}^2)$.
We assess the maximal direct CP violation in the above decays in the $SU(2)_L\times SU(2)_R\times U(1)_{B-L}$ model.
Additionally, we present a correlation between direct CP violation in these modes
 and one in $K\to\pi\pi$ decay parametrized by $\epsilon'$, since $W_R^+$ gauge boson
 has a sizable impact on the latter.}

\begin{document}

\maketitle
\flushbottom
\section{Introduction}

In the standard model (SM), direct CP violation in the Cabibbo-favored charmed meson decays
 is highly suppressed at the level of $\mathcal{O}(10^{-10})$~\cite{Delepine:2012xw}
 because no multiple tree and/or penguin diagrams with different CP phases can interfere.
Hence, if discovered, direct CP violation in these modes would immediately be a sign of new physics.
This is in contrast to the singly-Cabibbo-suppressed decays, where tree and penguin diagrams in the SM interfere to yield direct CP violation,
 and also $c\to s\bar{s}u$ and $c\to d\bar{d}u$ processes
 interfere through long-distance effects and may lead to sizable direct CP violation in the SM~\cite{Golden:1989qx}.
In this paper, we conduct a quantitative study on direct CP violation in Cabibbo-favored charmed meson decays with no final-state $K^0$, 
 namely, $D^0\to K^-\pi^+$, $D_s^+\to \eta\pi^+$ and $D_s^+\to \eta'\pi^+$ decays,
 in the $SU(2)_L\times SU(2)_R\times U(1)_{B-L}$ gauge extension of the SM.
Here, the absence of final-state $K^0$ ensures that
 Cabibbo-favored decay amplitudes do not interfere with doubly-Cabibbo-suppressed decay amplitudes via $K^0$-$\bar{K}^0$ mixing
 to induce SM contributions to direct CP violation~\cite{Bigi:1994aw,Xing:1995jg}.

In the $SU(2)_L\times SU(2)_R\times U(1)_{B-L}$ model, the right-right current operators, $(\bar{s}c)_{V+A}(\bar{u}d)_{V+A}$,
 coming from $W_R^+$ gauge boson exchange and the left-right current operators, $(\bar{s}c)_{V\pm A}(\bar{u}d)_{V\mp A}$,
 induced by $W_L^+$-$W_R^+$ mixing both contribute to the Cabibbo-favored decays.
However, the Wilson coefficients for the latter are suppressed by $\sim2m_b/m_t \simeq 1/20$ compared to the former
if the model naturally accommodates the bottom and top quark Yukawa couplings.
Therefore, we assume throughout this paper that the contribution of the right-right current operators 
 dominates over that of the left-right ones.
As a support for this assumption, we comment that the dominance of the right-right current contribution 
 has been observed in the study~\cite{Haba:2018byj} of direct CP violation in $K\to\pi\pi$ decay in the $SU(2)_L\times SU(2)_R\times U(1)_{B-L}$ model,
 where we have found that the right-right current contribution is larger by factor 5 than the left-right one,
 which indicates that although the hadronic matrix elements of the left-right operators are enhanced,
 this is insufficient to overcome the suppression of $2m_b/m_t \simeq 1/20$ on their Wilson coefficients.

The hadronic matrix elements of the right-right current operators are simply the minus of those of the left-left current operators, 
 due to parity symmetry of QCD.
Nevertheless, the decay amplitude from the right-right current operators and that from the left-left ones 
 acquire a non-trivial relative strong phase
 from difference in QCD corrections to the right-right and left-left current operators,
 which manifests itself as a difference between the ratio of the Wilson coefficients for 
 $(\bar{s}_\alpha c_\alpha)_{V+A}(\bar{u}_\beta d_\beta)_{V+A}$
 and $(\bar{s}_\alpha c_\beta)_{V+A}(\bar{u}_\beta d_\alpha)_{V+A}$ operators
 and the ratio of those for 
 $(\bar{s}_\alpha c_\alpha)_{V-A}(\bar{u}_\beta d_\beta)_{V-A}$
 and $(\bar{s}_\alpha c_\beta)_{V-A}(\bar{u}_\beta d_\alpha)_{V-A}$ operators
  ($\alpha,\beta$ denote color indices)
  at a given renormalization scale.
Ultimately, this difference is because the quark-gluon-quark-$W_L^+$($W_R^+$) box diagram in the fundamental theory
 contains terms proportional to $\log M_{W_L}^2$ ($\log M_{W_R}^2$),
 and hence the amount of QCD corrections to $W_L^+$ and $W_R^+$ gauge boson exchange diagrams 
 differ by $\sim(\alpha_s(M_{W_L}^2)/4\pi)\log(M_{W_R}^2/M_{W_L}^2)$.
Interestingly, this fact allows us to evaluate the relative strong phase,
 since the difference in QCD corrections at scales between $\mu\sim M_{W_R}$ and $\mu\sim M_{W_L}$
 is a short-distance effect.
Also, the scale-and-scheme-independent combinations~\cite{Buras:1998ra} of Wilson coefficients and hadronic matrix elements
 for $(\bar{s}_\alpha c_\alpha)_{V-A}(\bar{u}_\beta d_\beta)_{V-A}$ 
 and $(\bar{s}_\alpha c_\beta)_{V-A}(\bar{u}_\beta d_\alpha)_{V-A}$ operators can be estimated reliably
 with the diagrammatic approach with $SU(3)$ flavor symmetry~\cite{Zeppenfeld:1980ex,Chau:1982da,Gronau:1994bn,Hernandez:1994rh,Gronau:1994rj}, 
 which works successfully on the Cabibbo-favored charmed meson decays into two pseudoscalars~\cite{Bhattacharya:2009ps,Cheng:2010ry}.
 \footnote{
Earlier studies on the application of the diagrammatic approach with $SU(3)$ flavor symmetry 
 to charmed meson decays into two pseudoscalars are found in Refs.~\cite{Rosner:1999xd,Bhattacharya:2008ss}
 and in Refs.~\cite{Chau:1986du,Chau:1987tk,Chau:1989tk}.
}

Combining the strong phase thus evaluated and new CP-violating phases in the $SU(2)_L\times SU(2)_R\times U(1)_{B-L}$ model,
 we assess the maximal direct CP violation in $D^0\to K^-\pi^+$, $D_s^+\to \eta\pi^+$ and $D_s^+\to \eta'\pi^+$ decays.
Additionally, we investigate a correlation between direct CP violation in the above modes and one in $K \to \pi\pi$ decay parametrized by $\epsilon'$.
Previously, the authors have found~\cite{Haba:2018byj} that $W_R^+$ gauge boson in the $SU(2)_L\times SU(2)_R\times U(1)_{B-L}$ model with `charge symmetry'~\cite{Maiezza:2010ic}
has a sizable impact on $\epsilon'/\epsilon$ because $W_R^+$ exchange contributes to it at tree level.
It has been further revealed that
 the model with $\mathcal{O}(10)~\mathrm{TeV}$ $W_R^+$ boson mass
 can account for the incompatibility between the experimental data on $\epsilon'/\epsilon$~\cite{Batley:2002gn, AlaviHarati:2002ye, Abouzaid:2010ny}
 and the upper bound on $\epsilon'/\epsilon$~\cite{Buras:2015xba,Buras:2016fys}
 obtained with dual QCD approach and supported by
 lattice-based evaluations~\cite{Blum:2011ng, Blum:2015ywa, Bai:2015nea,Buras:2015yba, Kitahara:2016nld}.
\footnote{For other works on new physics contributions to $\epsilon'/\epsilon$, see Refs.~\cite{Cirigliano:2016yhc,Blanke:2015wba,Tanimoto:2016yfy, Kitahara:2016otd, 
 Endo:2016aws,Buras:2015jaq, Endo:2016tnu,Bobeth:2016llm,Buras:2015kwd,Chen:2018ytc,Chen:2018vog,Matsuzaki:2018jui}.}
Therefore, it is of particular interest how $\epsilon'/\epsilon$ and direct CP violation 
 in Cabibbo-favored charmed meson decays are correlated in the $SU(2)_L\times SU(2)_R\times U(1)_{B-L}$ model,
 and how the former constrains or predicts the latter.

This paper is organized as follows:
In Section 2, we briefly review the $SU(2)_L\times SU(2)_R\times U(1)_{B-L}$ gauge extension of the SM.
In Section 3, we give the effective Hamiltonian for the Cabibbo-favored charmed meson decays.
Section 4 presents our new results, where the diagrammatic amplitudes are reorganized in such a way
 that the decay amplitude coming from the right-right current operators are expressed in terms
 of the ratio of the Wilson coefficients that is calculable in short-distance QCD, and the diagrammatic amplitudes.
In Section 5, we show the results of our analysis on
 direct CP violation in $D^0\to K^-\pi^+$, $D_s^+\to \eta\pi^+$ and $D_s^+\to \eta'\pi^+$ decays,
 including its correlation with $\epsilon'/\epsilon$.
Section 6 summarizes the paper.
\\

\section{$SU(2)_L\times SU(2)_R\times U(1)_{B-L}$ model}

We briefly describe the $SU(2)_L\times SU(2)_R\times U(1)_{B-L}$ gauge extension of the SM.
Remind that charge symmetry~\cite{Maiezza:2010ic} is not imposed, unlike Ref.~\cite{Haba:2018byj}.
We summarize the matter content
in Table~\ref{content}.
\begin{table}[H]
\caption{Matter content and charge assignments
with $i$ being generation indices.}
\begin{center}
\begin{tabular}{|c|c|c|c|c|c|c|} \hline
Field               & Lorentz $SO(1,3)$ & $SU(3)_C$ & $SU(2)_L$  &  $SU(2)_R$  & $U(1)_{B-L}$ \\ \hline \hline
$Q_L^i$        & ({\bf 2},\,{\bf 1})  &  {\bf 3}        & {\bf 2}          &  {\bf 1}         &1/3                 \\ 
$Q_R^{c\,i}$& ({\bf 2},\,{\bf 1})  & $\bar{\bf 3}$   & {\bf 1}          &  {\bf 2}         &  $-1/3$           \\ 
$L_L^i$     & ({\bf 2},\,{\bf 1})  &{\bf 1}          & {\bf 2}          &  {\bf 1}         &  $-1$               \\
$L_R^{c\,i}$& ({\bf 2},\,{\bf 1}) &{\bf 1}        & {\bf 1}          &  {\bf 2}         &  $1$          \\ \hline
$\Phi$                & {\bf 1}                   &{\bf 1}        &  {\bf 2}         &  {\bf 2}          &  0      \\
$\Delta_L$      & {\bf 1}                     &{\bf 1}       &  {\bf 3}          &  {\bf 1}         & $2$ \\ 
$\Delta_R$      & {\bf 1}                     &{\bf 1}       &  {\bf 1}          &  {\bf 3}         & $-2$ \\ \hline 
\end{tabular}
\end{center}
\label{content}
\end{table}
\noindent
$SU(2)_L\times SU(2)_R\times U(1)_{B-L}$ gauge interactions and Yukawa interactions of quarks are described as
\bea
-{\cal L}&\supset& Q_L^{i\,\dagger} \, \bar{\sigma}_\mu\left( \frac{1}{2}g_L \sigma^a W_L^{a\,\mu}+\frac{1}{3}g_X X^\mu\right)Q_L^i
+Q_R^{c\,i\,\dagger} \, \bar{\sigma}_\mu\left( -\frac{1}{2}g_R (\sigma^a)^T W_R^{a\,\mu}-\frac{1}{3}g_X X^\mu\right)Q_R^{c\,i}
\nonumber\\
&+&(Y_q)_{ij}\,Q_L^{i\,\dagger}\Phi\epsilon_s(Q_R^{c\,j})^*+(\tilde{Y}_q)_{ij}\,Q_L^{i\,\dagger}(\epsilon_g^T\Phi^*\epsilon_g)\epsilon_s(Q_R^{c\,j})^*+{\rm H.c.}\label{lag}
\eea
$\Delta_R$ acquires a VEV, $v_R$, to break $SU(2)_R\times U(1)_{B-L} \to U(1)_Y$, 
and $\Phi$ further gains a VEV, $\langle\Phi\rangle={\rm diag}(v\sin\beta, \ v\cos\beta e^{i\,\alpha})$ 
with $v\simeq246$~GeV, to trigger the electroweak symmetry breaking.
As a result, the charged $SU(2)_L$ gauge boson, $W_L^+$, 
 and the charged $SU(2)_R$ gauge boson, $W_R^+$, 
 mix and form two mass eigenstates $W^+$, $W'^+$ as
\bea
-{\cal L}&\supset& M_W^2 W^+ W^- + M_{W'}^2 {W'}^+ {W'}^-,
\ \ \ \ \ \begin{pmatrix} 
      W_L^+ \\
      W_R^+ \\
   \end{pmatrix}=\begin{pmatrix} 
         \cos\zeta &-e^{-i\,\alpha}\sin\zeta  \\
         e^{i\,\alpha}\sin\zeta  & \cos\zeta \\
      \end{pmatrix}
         \begin{pmatrix} 
            W^+ \\
            W'^+ \\
         \end{pmatrix},\label{wlwr}\\
\sin\zeta&\simeq&\frac{g_R}{g_L}\frac{M_{W}^2}{M_{W'}^2}\sin(2\beta) \ \ \ \ \ \mathrm{for} \ M_{W'}^2\gg M_W^2.
\label{sinzeta}
\eea
The up-type quark mass matrix, $M_u$, and the down-type one, $M_d$, are given by
\footnote{
$U_R^i\equiv \epsilon_s (U_R^{c\,i})^*$, $D_R^i\equiv \epsilon_s (D_R^{c\,i})^*$.
}
\begin{align} 
M_u&=\frac{v}{\sqrt{2}}\left(\sin\beta Y_q+\cos\beta e^{-i\,\alpha}\tilde{Y}_q\right), \ \ \
M_d=\frac{v}{\sqrt{2}}\left(\cos\beta e^{i\,\alpha}Y_q+\sin\beta\tilde{Y}_q\right),
\label{quarkmass}
\end{align}
 which are diagonalized as
 $M_u=V_{uL}^\dagger$diag$(m_u,\,m_c,\,m_t) V_{uR}$ and $M_d=V_{dL}^\dagger$diag$(m_d,\,m_s,\,m_b) V_{dR}$
 with unitary matrices $V_{uL},V_{uR},V_{dL},V_{dR}$.
Then, we obtain the SM Cabibbo-Kobayashi-Maskawa matrix as $V_L=V_{uL}V_{dL}^\dagger$, 
 and the corresponding flavor mixing matrix for right-handed quarks as $V_R=V_{uR}V_{dR}^\dagger$.
From Eq.~(\ref{wlwr}), 
 we find that the charged-current interactions are described by the following term in the unitary gauge:
\begin{align} 
-{\cal L}&\supset\frac{1}{\sqrt{2}} \, \bar{U}^i \, W^{+\,\mu}\gamma_\mu
\left\{g_L(V_L)_{ij}\cos\zeta P_L+g_R(V_R)_{ij} \, e^{i\,\alpha}\sin\zeta P_R\right\}\,D^j
\nonumber \\
&+\frac{1}{\sqrt{2}} \, \bar{U}^i \, W'^{+\,\mu} \gamma_\mu
\left\{-g_L(V_L)_{ij} \, e^{-i\,\alpha}\sin\zeta P_L+g_R(V_R)_{ij}\cos\zeta P_R\right\}\,D^j+{\rm H.c.}
\label{deltaf=1}
\end{align}
From Eq.~(\ref{quarkmass}), it is clear that the top and bottom Yukawa couplings are derived without fine-tuning 
 only when $\tan\beta\simeq m_b/m_t$ holds,
 which, combined with Eq.~(\ref{sinzeta}), gives $\sin\zeta\simeq (2m_b/m_t)(g_R/g_L)(M_W^2/M_{W'}^2)$.
Then, one finds from Eq.~(\ref{deltaf=1}) that 
 the Wilson coefficients for the left-right currents obtained by integrating out $W^+$ are suppressed by $2m_b/m_t$ 
 compared to those for the right-right currents obtained by integrating out $W'^+$.
\\

\section{Effective Hamiltonian for Cabibbo-favored $\Delta C=1$ process}

The effective Hamiltonian for Cabibbo-favored $\Delta C=1$ process reads,
\bea
\mathcal{H}^{\Delta C=1}_{\mathrm{eff}}=
\displaystyle\sum_{i=1}^2(C_i^{\mathrm{LL}}Q_{i}^\mathrm{LL}+C_i^{\RR}Q_{i}^{\RR}
).\label{Eq:CAH}\eea
The operators above are defined as
\bea
Q_{1}^{\mathrm{LL}}=(\bar{s}_\alpha c_\beta)_{V-A}
(\bar{u}_\beta d_\alpha)_{V-A},\quad
Q_{2}^{\mathrm{LL}}=(\bar{s}_\alpha c_\alpha)_{V-A}
(\bar{u}_\beta d_\beta)_{V-A},
\nonumber\\
Q_{1}^{\mathrm{RR}}=(\bar{s}_\alpha c_\beta)_{V+A}
(\bar{u}_\beta d_\alpha)_{V+A},\quad
Q_{2}^{\mathrm{RR}}=(\bar{s}_\alpha c_\alpha)_{V+A}
(\bar{u}_\beta d_\beta)_{V+A},
\label{Eq:CAop}
\eea
where $(\bar{q}q^\prime)_{V-A}$ and $(\bar{q}q^\prime)_{V+A}$ stand for
 $\bar{q}\gamma_\mu (1-\gamma_5)q^\prime$ and $\bar{q}\gamma_\mu (1+\gamma_5)q^\prime$, respectively,
 and $\alpha,\beta$ denote QCD color indices.
$C_{i}^{\mathrm{LL}} (i=1, 2)$ in Eq.~(\ref{Eq:CAH})
represents the SM contribution,
while $C_{i}^{\mathrm{RR}}$ arises from $W_R^+$ gauge boson exchange.
In this paper, we neglect the left-right current operators
$(\bar{s}c)_{V\pm A} (\bar{u}d)_{V\mp A}$ induced by $W_L^+$-$W_R^+$ mixing,
 because the corresponding Wilson coefficients are suppressed by $2m_b/m_t\simeq1/20$ compared to $C_{i}^{\mathrm{RR}}$
 if there is no fine-tuning in deriving the bottom quark Yukawa coupling.
\par
The renormalization group equation (RGE) of the Wilson
coefficients is divided into two pieces for chirality-flipped sectors.
At leading order, it reads
\bea
\mu\frac{\mathrm{d}}{\mathrm{d}\mu}\vec{C}_{\mathrm{LL}}
=\gamma^\mathrm{T}\vec{C}_{\mathrm{LL}},\qquad
\mu\frac{\mathrm{d}}{\mathrm{d}\mu}\vec{C}_{\mathrm{RR}}
=\gamma^\mathrm{T}\vec{C}_{\mathrm{RR}},\qquad
\gamma=\begin{pmatrix}
-2 & 6\\
6 & -2
\end{pmatrix}\frac{\alpha_s}{4\pi},
\label{Eq:Ren}
\eea
where $\vec{C}_\mathrm{LL}=(C_1^{\mathrm{LL}}, C_2^{\mathrm{LL}})^\mathrm{T}$,
 $\vec{C}_\mathrm{RR}=(C_1^{\mathrm{RR}}, C_2^{\mathrm{RR}})^\mathrm{T}$,
 and the anomalous dimension matrix $\gamma$ is common for LL and RR sectors.
The initial conditions for the RGE at leading order are
\bea
C_1^{\mathrm{LL}}(\mu_{W})&=&0,\\
C_2^{\mathrm{LL}}(\mu_{W})&=&
\frac{G_F}{\sqrt{2}}V_{cs}^{\mathrm{L}*}V_{ud}^\mathrm{L},\\
C_1^{\mathrm{RR}}(\mu_{W^\prime})&=&0,\\
C_2^{\mathrm{RR}}(\mu_{W^\prime})&=&
\frac{G_F}{\sqrt{2}}V_{cs}^{\mathrm{R}*}V_{ud}^\mathrm{R}
\left(\frac{g_R}{g_L}\frac{M_W}{M_{W'}}\right)^2,
\label{Eq:73}
\eea
with $\mu_W\sim M_W, \mu_{W'}\sim M_{W'}$.
The RGE (\ref{Eq:Ren}) is diagonalized in the basis of
$C^{\mathrm{LL}}_\pm=C_1^{\mathrm{LL}}\pm C_2^{\mathrm{LL}}$
and $C^{\mathrm{RR}}_\pm=C_1^{\mathrm{RR}}\pm C_2^{\mathrm{RR}}$
so that the RG evolution is simply described without operator mixing.

\section{Decay amplitudes from right-right current operators}

Hereafter, we exclusively work under the assumption of $SU(3)$ flavor symmetry of $u,d,s$ quarks.
The amplitudes of charmed meson decays to two pseudoscalars ($D\to PP$)
 can be categorized by diagrammatic topologies~\cite{Chau:1986du,Chau:1987tk,Chau:1989tk,Rosner:1999xd,Bhattacharya:2008ss,Bhattacharya:2009ps,Cheng:2010ry}.
For the Cabibbo-favored $D\to PP$ decays,
 the diagrammatic amplitudes consist of $T$(tree), $C$(color-suppressed tree), 
 $A$(annihilation) and $E$(exchange) diagrams.
In addition, Ref.~\cite{Buras:1998ra} has clarified the correspondence 
 between the diagrammatic amplitudes and the scale-and-scheme-independent combinations 
 of Wilson coefficients and operators.

For the left-left and right-right current contributions, the diagrammatic amplitudes are rewritten as
\bea
T_{\LL}=C_1^{\LL}(\mu)\braket{Q_1(\mu)}_{\mathrm{CE}}
+C_2^{\LL}(\mu) \braket{Q_2(\mu)}_{\mathrm{DE}}, \ \ \
T_{\RR}=-C_1^{\RR}(\mu)\braket{Q_1(\mu)}_{\mathrm{CE}}
-C_2^{\RR}(\mu) \braket{Q_2(\mu)}_{\mathrm{DE}},
\label{Eq:T}\qquad\\
C_{\LL}=C_1^{\LL}(\mu) \braket{Q_1(\mu)}_{\mathrm{DE}}
+C_2^{\LL}(\mu) \braket{Q_2(\mu)}_{\mathrm{CE}}, \ \ \
C_{\RR}=-C_1^{\RR}(\mu) \braket{Q_1(\mu)}_{\mathrm{DE}}
-C_2^{\RR}(\mu) \braket{Q_2(\mu)}_{\mathrm{CE}},\qquad\\
A_{\LL}=C_1^{\LL}(\mu)\braket{Q_1(\mu)}_{\mathrm{CA}}
+C_2^{\LL}(\mu) \braket{Q_2(\mu)}_{\mathrm{DA}}, \ \ \
A_{\RR}=-C_1^{\RR}(\mu)\braket{Q_1(\mu)}_{\mathrm{CA}}
-C_2^{\RR}(\mu) \braket{Q_2(\mu)}_{\mathrm{DA}},\qquad\\
E_{\LL}=C_1^{\LL}(\mu) \braket{Q_1(\mu)}_{\mathrm{DA}}
+C_2^{\LL}(\mu) \braket{Q_2(\mu)}_{\mathrm{CA}}, \ \ \ 
E_{\RR}=-C_1^{\RR}(\mu) \braket{Q_1(\mu)}_{\mathrm{DA}}
-C_2^{\RR}(\mu) \braket{Q_2(\mu)}_{\mathrm{CA}},\qquad
\label{Eq:A}
\eea
 where $\mu$ denotes a common renormalization scale for the Wilson coefficients and operators of both left-left and right-right currents.
 $\braket{Q_i(\mu)}$ denotes a hadronic matrix element defined by
 $\braket{Q_i(\mu)}=\bra{PP}Q^{\LL}_i(\mu)\ket{D}$, whose subscript represents the connected emission (CE),
 the disconnected emission (DE), the connected annihilation (CA) and the disconnected annihilation (DA), respectively~\cite{Buras:1998ra}.
We have used $\bra{PP}Q^{\LL}_i(\mu)\ket{D}=-\bra{PP}Q^{\RR}_i(\mu)\ket{D} (i=1, 2)$ 
 that follows from parity conservation of QCD.
We emphasize that each of
$T_{\LL},T_{\RR},C_{\LL},\\
C_{\RR},A_{\LL},A_{\RR},E_{\LL},E_{\RR}$ 
 is independent of renormalization scale and scheme~\cite{Buras:1998ra}.
\par
By rewriting $Q_1^{\mathrm{LL}}$ as 
 $(\bar{s}_\alpha d_\alpha)_{V-A}(\bar{u}_\beta c_\beta)_{V-A}$ through the Fierz rearrangement,
 we obtain the following relations based on $SU(3)$ flavor symmetry of $u,d,s$ quarks:
\bea
\braket{Q_1(\mu)}_{\mathrm{CE}}&=&\braket{Q_2(\mu)}_{\mathrm{CE}},\label{ce}\\
\braket{Q_1(\mu)}_{\mathrm{DE}}&=&\braket{Q_2(\mu)}_{\mathrm{DE}},\label{de}\\
\braket{Q_1(\mu)}_{\mathrm{CA}}&=&\braket{Q_2(\mu)}_{\mathrm{CA}},\label{ca}\\
\braket{Q_1(\mu)}_{\mathrm{DA}}&=&\braket{Q_2(\mu)}_{\mathrm{DA}}\label{da}.
\eea
Henceforth, the subscripts of the operators are omitted.
Using Eqs.~(\ref{ce}--\ref{da}), we can re-express the diagrammatic amplitudes
 in terms of $C^{\mathrm{LL}}_\pm=C_1^{\mathrm{LL}}\pm C_2^{\mathrm{LL}}$ as
\bea
T_{\LL}&=&
C_+^{\LL}\frac{\braket{Q}_{\mathrm{CE}}
+\braket{Q}_{\mathrm{DE}}}{2}+C_-^{\LL}\frac{\braket{Q}_{\mathrm{CE}}
-\braket{Q}_{\mathrm{DE}}}{2},
\\
C_{\LL}&=&
C_+^{\LL}\frac{\braket{Q}_{\mathrm{CE}}
+\braket{Q}_{\mathrm{DE}}}{2}-C_-^{\LL}\frac{\braket{Q}_{\mathrm{CE}}
-\braket{Q}_{\mathrm{DE}}}{2},
\\
A_{\LL}&=&
C_+^{\LL}\frac{\braket{Q}_{\mathrm{CA}}
+\braket{Q}_{\mathrm{DA}}}{2}+C_-^{\LL}\frac{\braket{Q}_{\mathrm{CA}}
-\braket{Q}_{\mathrm{DA}}}{2},
\\
E_{\LL}&=&
C_+^{\LL}\frac{\braket{Q}_{\mathrm{CA}}
+\braket{Q}_{\mathrm{DA}}}{2}-C_-^{\LL}\frac{\braket{Q}_{\mathrm{CA}}
-\braket{Q}_{\mathrm{DA}}}{2}.
\eea
It follows that the right-right current contributions can be rewritten as
\bea
T_{\RR}&=&-\frac{C_{+}^{\RR}}{C_{+}^{\LL}}\frac{T_{\LL}+C_{\LL}}{2}-\frac{C_{-}^{\RR}}{C_{-}^{\LL}}\frac{T_{\LL}-C_{\LL}}{2},\label{trr}\\
C_{\RR}&=&-\frac{C_{+}^{\RR}}{C_{+}^{\LL}}\frac{T_{\LL}+C_{\LL}}{2}+\frac{C_{-}^{\RR}}{C_{-}^{\LL}}\frac{T_{\LL}-C_{\LL}}{2},\\
A_{\RR}&=&-\frac{C_{+}^{\RR}}{C_{+}^{\LL}}\frac{A_{\LL}+E_{\LL}}{2}-\frac{C_{-}^{\RR}}{C_{-}^{\LL}}\frac{A_{\LL}-E_{\LL}}{2},\\
E_{\RR}&=&-\frac{C_{+}^{\RR}}{C_{+}^{\LL}}\frac{A_{\LL}+E_{\LL}}{2}+\frac{C_{-}^{\RR}}{C_{-}^{\LL}}\frac{A_{\LL}-E_{\LL}}{2}.\label{err}
\eea
The ratio of the Wilson coefficients, $C_\pm^{\mathrm{RR}}/C_\pm^{\mathrm{LL}}$, in Eqs.~(\ref{trr}--\ref{err}) is
 independent of renormalization scale and scheme.
As a reference, we find, at the leading order,
\bea
\frac{C_\pm^{\mathrm{RR}}(\mu)}{C_\pm^{\mathrm{LL}}(\mu)}
=(\eta_{\mu_{W^\prime}}^{\mu_{W}})^{-\frac{\lambda_{0\pm}}{2\beta_0}}
\left(\frac{g_R}{g_L}\frac{M_W}{M_{W^\prime}}\right)^2
\frac{V_{cs}^{\mathrm{R}*}V_{ud}^{\mathrm{R}}}{V_{cs}^{\mathrm{L}*}V_{ud}^{\mathrm{L}}},\label{wilsonratio}
\eea
 where $\lambda_{0+}=4$, $\lambda_{0-}=-8$, and $\beta_0=11-2n_f/3$ with $n_f=6$, and
we have defined the QCD correction factor as $\eta^{\mu_1}_{\mu_2}=\alpha_s(\mu_1)/\alpha_s(\mu_2)$.
The next-leading order (NLO) QCD corrections to
Eq.~(\ref{wilsonratio}) are found in Eq.~(\ref{Eq:ratio}).

The diagrammatic amplitudes have been determined through
 a phenomenological fitting of $D\to PP$ decay partial widths
 in Ref.~\cite{Bhattacharya:2009ps} (see also Ref.~\cite{Cheng:2010ry}).
In that study, an important assumption is
that OZI-suppressed diagrams for $D^0\to \bar{K}^0\eta$,
$D^0\to \bar{K}^0\eta'$, $D_s^+\to \pi^+\eta$,
$D_s^+\to \pi^+\eta'$ decays
 are negligible in the partial widths.
Also, the $SU(3)$ flavor symmetry is assumed.
These assumptions are justified for the Cabibbo-favored decays,
 since a good fit with $\chi^2=1.79$ for 1 degree of freedom
 for fixed $\eta-\eta^\prime$ mixing angle is obtained in that study.
\footnote{
A better fit has been found in Ref.~\cite{Li:2012cfa},
 where factorization-assisted topological-amplitude approach
 with the inclusion of SU(3) breaking effects is used.
}
In this paper, we employ the result of Ref.~\cite{Bhattacharya:2009ps}
 by fixing the $\eta-\eta^\prime$ mixing angle at $19.5^\circ$.
Assuming that
the contributions of the right-right current operators to the partial widths
are negligible, one finds~\cite{Bhattacharya:2009ps}
(in $10^{-6}\ \mathrm{GeV}$ unit),
$T_{\mathrm{LL}}=2.927\pm 0.022,
C_{\mathrm{LL}}=(2.337\pm0.027)\: \mathrm{exp}[i(-151.66\pm0.63)^\circ],
A_{\mathrm{LL}}=(0.33\pm0.14)\: \mathrm{exp}[i(70.47\pm10.90)^\circ]$ and
$E_{\mathrm{LL}}=(1.573\pm0.032)\: \mathrm{exp}[i(120.56\pm1.03)^\circ]$.

\section{Numerical analysis on direct CP violation}

In the SM, direct CP violation in the Cabibbo-favored decays is generated via
 the interference between the tree diagram
 and the box and di-penguin diagrams~\cite{Delepine:2012xw}.
CP asymmetry in $D^0\to K^-\pi^+$ decay rate is estimated to be $1.4\times10^{-10}$
 in Ref.~\cite{Delepine:2012xw}.
We infer that direct CP violation is suppressed similarly in all Cabibbo-favored modes,
 and therefore neglect the SM contribution in all modes.
Provided the contribution of the right-right current is small,
 CP asymmetry in the decay rates can be expanded as
\bea
A_{\mathrm{CP}}^{D\to f}
&=&\frac{\Gamma[D\to f]-\Gamma[\bar{D}\to \bar{f}]}
{\Gamma[D\to f]+\Gamma[\bar{D}\to \bar{f}]}
\simeq\mathrm{Re}\left[\frac{({\cal A}_{f})_\RR}{({\cal A}_{f})_\LL}
-\frac{(\bar{{\cal A}}_{\bar{f}})_\RR}{(\bar{{\cal A}}_{\bar{f}})_\LL}\right].\label{Eq:ASYEXP}
\eea
The diagrammatic amplitude of each Cabibbo-favored decay is given in Tab.~\ref{Tab:1}.
By using the relations Eqs.~(\ref{trr}-\ref{err}) and the leading order expression for the Wilson coefficient ratio Eq.~(\ref{wilsonratio}),
 we find that the asymmetry takes a simple form,
\bea
A_{\mathrm{CP}}^{D\to f}=F_{\mathrm{CP}}^{D\to f}
\left[
\left(\eta^{\mu_W}_{\mu_{W'}}\right)^{-\frac{2}{7}}
-\left(\eta^{\mu_W}_{\mu_{W'}}\right)^{\frac{4}{7}}
\right]\left(\frac{g_R}{g_L}\frac{M_W}{M_{W'}}\right)^2
\mathrm{Im}\left(\frac{V_{cs}^{\mathrm{R}*}V_{ud}^{\mathrm{R}}}
{V_{cs}^{\mathrm{L}*}V_{ud}^{\mathrm{L}}}
\right),\label{Eq:ASS}
\eea
 where $F_{\mathrm{CP}}^{D\to f}$ is a process-dependent factor,
which is summarized in Tab.~\ref{Tab:1}.
The QCD correction factor and CP phase dependence in Eq.~(\ref{Eq:ASS})
 are common for all Cabibbo-favored modes.
Note that $A_{\mathrm{CP}}^{D^0\to \bar{K^0}\pi^+}$
 vanishes because $T_{\mathrm{RR}}+C_{\mathrm{RR}}$ and $T_{\mathrm{LL}}+C_{\mathrm{LL}}$
 have an identical strong phase.
In Appendix~A,
NLO QCD corrections
with the appropriate threshold corrections
at the matching scales $\mu_{W}$ and $\mu_{W^\prime}$, which we use in
the numerical analysis, are given.

\begin{figure}[H]
  \begin{center}
    \includegraphics[width=12cm]{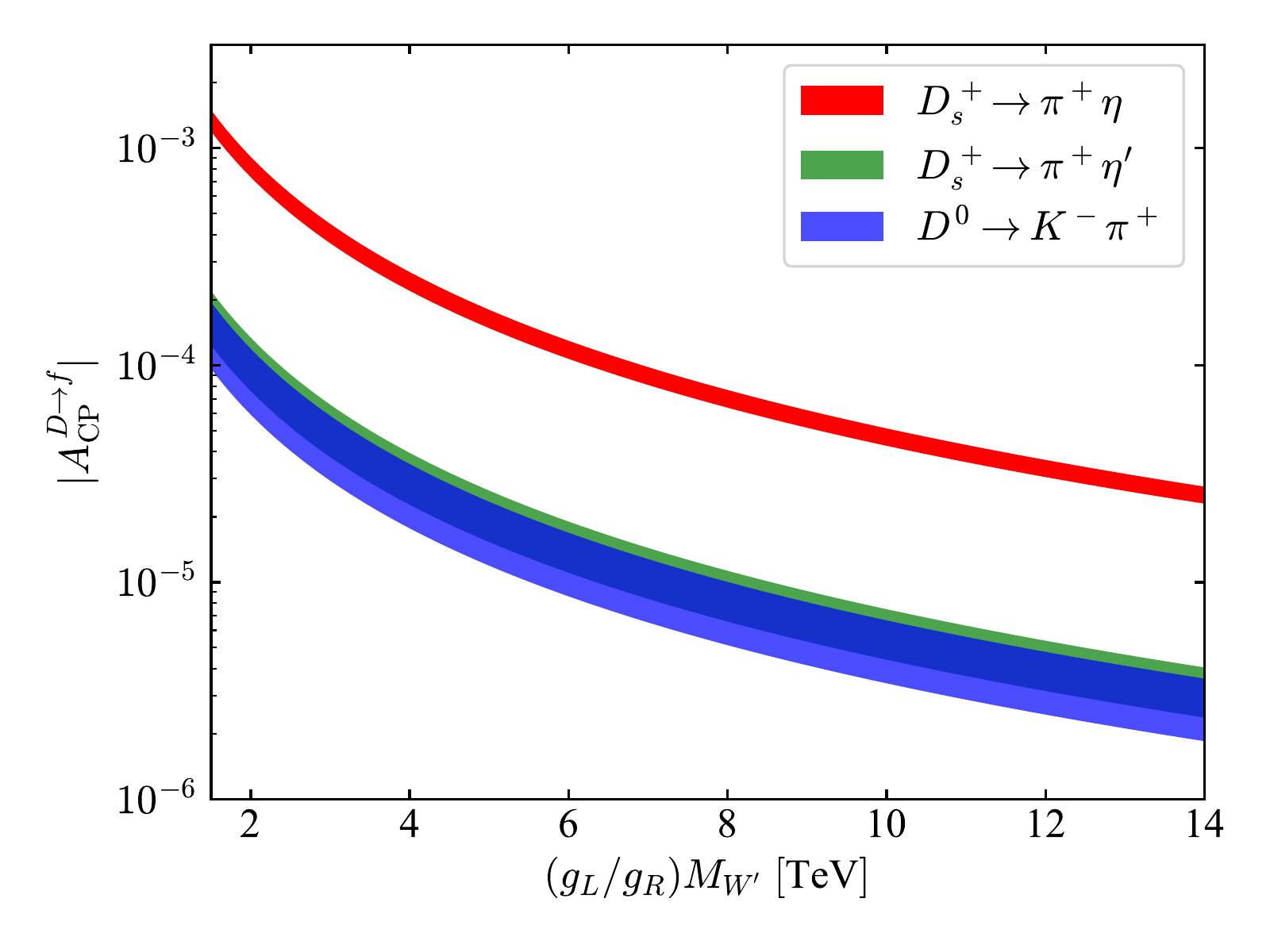}
    \caption{Absolute value of the maximal CP asymmetry of the partial width of Cabibbo-favored charmed meson decays
    in the $SU(2)_L\times SU(2)_R\times U(1)_{B-L}$ model (without charge symmetry).
    The bands represent the combination of theoretical uncertainty evaluated by varying the matching scales as
    $M_W/2\leq\mu_W\leq 2M_W$ and $M_{W'}/2\leq\mu_{W'}\leq 2M_{W'}$,
     and uncertainty from
    the $1\sigma$ errors of the diagrammatic amplitudes.}
    \label{Fig:1}
  \end{center}
\end{figure}

In Fig.~\ref{Fig:1}, maximal CP asymmetries
in $D^0\to K^-\pi^+$, $D^+_s\to \pi^+\eta$ and
$D^+_s\to \pi^+\eta^\prime$ in the $SU(2)_L\times SU(2)_R\times U(1)_{B-L}$ are plotted
by taking
 $\mathrm{Im}\left(V_{cs}^{\mathrm{R}*}V_{ud}^{\mathrm{R}}/
V_{cs}^{\mathrm{L}*}V_{ud}^{\mathrm{L}}\right)=1/\cos^2\theta_C$
 ($\theta_C$ denotes the SM Cabibbo angle).
To estimate theoretical uncertainty, we have varied the matching scales $\mu_W$ and $\mu_{W'}$ 
 in the range $M_W/2\leq\mu_W\leq 2M_W$ and $M_{W'}/2\leq\mu_{W'}\leq 2M_{W'}$, respectively.
Also, the $1\sigma$ errors of the diagrammatic amplitudes in Ref.~\cite{Bhattacharya:2009ps} are considered as a source of uncertainty.
We observe in Fig.~\ref{Fig:1} that the asymmetry is specially enhanced in $D^+_s\to \pi^+\eta$ decay,
 due to the relatively large process-dependent factor.
Note that we do not study the other Cabibbo-favored decays,
 because they include a final-state $\bar{K}^0$ and are thus
 observed via $K^0$-$\bar{K}^0$ mixing.
Hence, the amplitudes of Cabibbo-favored and doubly-Cabibbo-suppressed decays
 interfere to yield non-negligible CP asymmetry in the SM.

\begin{figure}[H]
  \begin{center}
    \includegraphics[width=12cm]{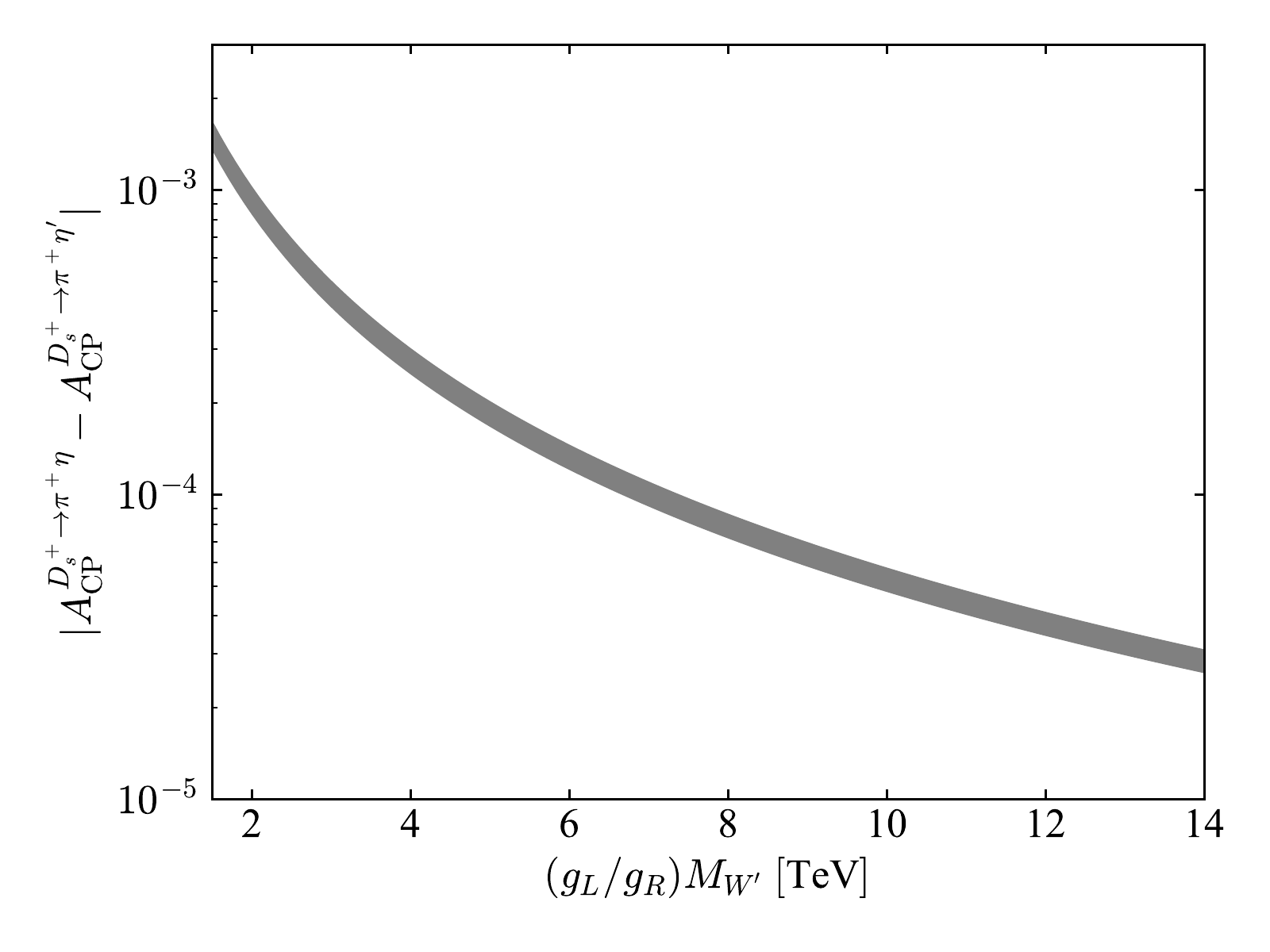}
    \caption{Maximal difference in the CP asymmetry in 
    $D^+_s\to \pi^+\eta$ and
    $D^+_s\to \pi^+\eta'$ in the $SU(2)_L\times SU(2)_R\times U(1)_{B-L}$ model (without charge symmetry).
    The bands represent the combination of theoretical uncertainty evaluated by varying the matching scales as
     $M_W/2\leq\mu_W\leq 2M_W$ and $M_{W'}/2\leq\mu_{W'}\leq 2M_{W'}$,
     and uncertainty from
     the $1\sigma$ errors of the diagrammatic amplitudes.}
    \label{Fig:2}
  \end{center}
\end{figure}

In real experiments, one measures the difference of the CP asymmetries in two processes,
 to nullify asymmetry in the production cross sections at $pp$ colliders or 
 a slight asymmetry in the production kinematics at $e^+e^-$ colliders (due to $Z$-photon interference),
 and asymmetry in the efficiency of charged meson detection.
Consequently, most of the systematic uncertainties cancel.
For the search for direct CP violation in Cabibbo-favored decays in the $SU(2)_L\times SU(2)_R\times U(1)_{B-L}$ model, 
 we suggest that one measure
\bea
A_{\mathrm{CP}}^{D_s^+\to \pi^+\eta}-A_{\mathrm{CP}}^{D_s^+\to \pi^+\eta'},\label{acpdiff}
\eea
 because the two asymmetries are predicted to have opposite signs in Tab.~\ref{Tab:1}
 (note the signs of $F_{\mathrm{CP}}^{D\to f}$) and $A_{\mathrm{CP}}^{D_s^+\to \pi^+\eta}$ is sizable.
Also, asymmetries in the $D_s^{\pm}$ production and the $\pi^\pm$ detection efficiency largely cancel between the two processes.
In Fig.~\ref{Fig:2}, we plot the maximal difference in
 the CP asymmetries in $D^+_s\to\pi^+\eta$ and $D^+_s\to\pi^+\eta'$
 by again taking
 $\mathrm{Im}\left(V_{cs}^{\mathrm{R}*}V_{ud}^{\mathrm{R}}/
V_{cs}^{\mathrm{L}*}V_{ud}^{\mathrm{L}}\right)=1/\cos^2\theta_C$.
We comment that, as shown in Appendix~B, 
 our prediction for the CP asymmetry difference Eq.~(\ref{acpdiff}), 
 which has been derived by assuming $SU(3)$ flavor symmetry,
 is not much affected by $SU(3)$ flavor symmetry breaking.

We make a crude estimate on the statistical uncertainty in a measurement of Eq.~(\ref{acpdiff})
 at Belle II with 50~ab$^{-1}$ of data.
Reference~\cite{Won:2011ku} reports that with 791~fb$^{-1}$ of data at Belle,
 statistical uncertainty of the CP asymmetry in the number of reconstructed events 
 $(N_{\mathrm{rec}}(D\to f)-N_{\mathrm{rec}}(\bar{D}\to \bar{f}))/(N_{\mathrm{rec}}(D\to f)+N_{\mathrm{rec}}(\bar{D}\to \bar{f}))$ 
 is 1.13\% for $D^+\to\pi^+\eta$ and 1.12\% for $D^+\to\pi^+\eta'$.
Assuming that the signal efficiencies (1.6\%-1.7\%) are the same for $D^+\to\pi^+\eta(')$ and $D_s^+\to\pi^+\eta(')$,
 and using the branching ratios found in Ref.~\cite{Patrignani:2016xqp},
 we estimate the statistical uncertainty at Belle II with 50~ab$^{-1}$ of data to be
 $\Delta(A_{\mathrm{CP}}^{D_s^+\to \pi^+\eta}-A_{\mathrm{CP}}^{D_s^+\to \pi^+\eta'})=$0.08\%.
Next, we estimate the statistical uncertainty in a measurement of Eq.~(\ref{acpdiff}) at LHCb with 50~fb$^{-1}$ of data.
Reference~\cite{Aaij:2017eux} reports that with 1~fb$^{-1}$ of data at 7~TeV and 2~fb$^{-1}$ of data at 8~TeV LHCb,
 the signal yield of $D_s^\pm\to\pi^\pm\eta'$ processes is 152$\times10^3$.
Making a rough approximation that the signal yield with 2~fb$^{-1}$ of data at 8~TeV is twice the yield with 1~fb$^{-1}$ of data at 7~TeV,
 and performing a na\"ive rescaling of the number of events by $\times 200$ based on Ref.~\cite{Bediaga:2012py},
 the signal yield of $D_s^\pm\to\pi^\pm\eta'$ processes with 50~fb$^{-1}$ of data is estimated to be 10$^7$.
Further assuming that the signal efficiencies for $D_s^\pm\to\pi^\pm\eta'$ and $D_s^\pm\to\pi^\pm\eta$ are the same,
 the statistical uncertainty with 50~fb$^{-1}$ of data is found to be
 $\Delta(A_{\mathrm{CP}}^{D_s^+\to \pi^+\eta}-A_{\mathrm{CP}}^{D_s^+\to \pi^+\eta'})=$0.06\%.
We find that if the $SU(2)_R$ gauge coupling is enhanced as $g_R=2g_L$, 
 one may hope to discover direct CP violation in Cabibbo-favored decays even with $M_{W'}=4$~TeV 
 (this parameter point is nearly consistent with
 the bound on $Z'$ derived in Refs.~\cite{Patra:2015bga,Lindner:2016lpp}).

\begin{figure}[H]
  \begin{center}
    \includegraphics[width=12cm]{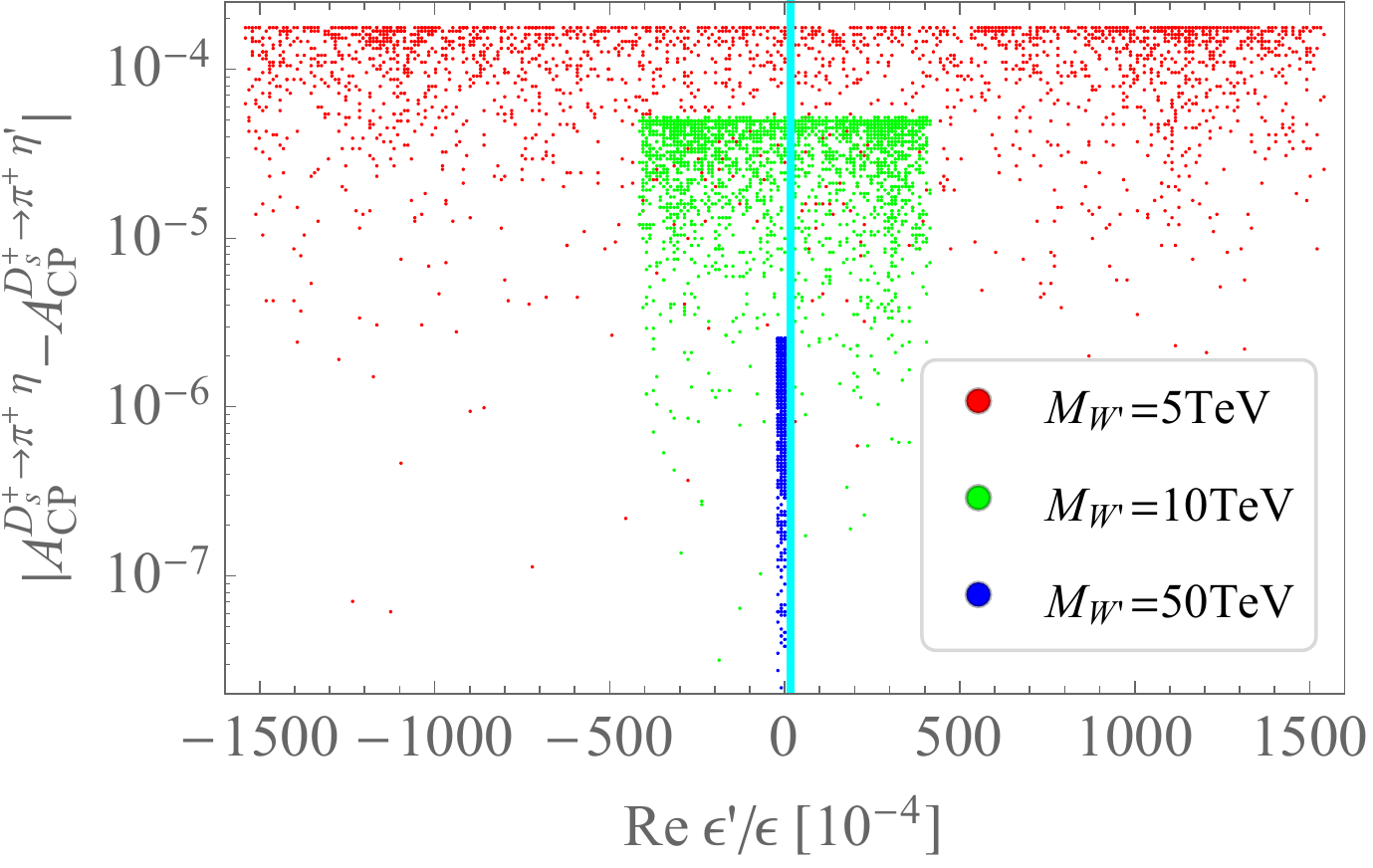}
    \caption{Correlated prediction for difference in the CP asymmetry in 
    $D^+_s\to \pi^+\eta$ and
    $D^+_s\to \pi^+\eta'$,
     and $\mathrm{Re}(\epsilon'/\epsilon)$, in the $SU(2)_L\times SU(2)_R\times U(1)_{B-L}$ model with charge symmetry.
    The red, green and blue dots represent the parameter points with randomly generated values of new CP violating phases
     for $M_{W'}=5$~TeV, 10~TeV and 50~TeV, respectively.    
    The cyan band stands for the $1\sigma$ range of
    the experimental value of $\mathrm{Re}(\epsilon'/\epsilon)$ \cite{Patrignani:2016xqp}.}
    \label{Fig:3}
  \end{center}
\end{figure}

In Fig.~\ref{Fig:3}, a correlated prediction for
 the CP asymmetry difference $A_{\mathrm{CP}}^{D_s^+\to \pi^+\eta}-A_{\mathrm{CP}}^{D_s^+\to \pi^+\eta'}$
 and $\mathrm{Re}(\epsilon'/\epsilon)$ calculated in Ref.~\cite{Haba:2018byj}
 in the $SU(2)_L\times SU(2)_R\times U(1)_{B-L}$ model
 is presented.
Here, as with Ref.~\cite{Haba:2018byj}, we impose `charge symmetry'~\cite{Maiezza:2010ic} on the model,
 which gives $g_L=g_R$ and $V_{ud}^{\mathrm{R}}=(V_{ud}^{\mathrm{L}})^*e^{-i\,\psi_d}$,
 $V_{cs}^{\mathrm{R}}=(V_{cs}^{\mathrm{L}})^*e^{i(\phi_c-\psi_s)}$,
 $V_{us}^{\mathrm{R}}=(V_{us}^{\mathrm{L}})^*e^{-i\,\psi_s}$
 with $\psi_d,\psi_s,\phi_c$ being arbitrary CP-violating phases.
We thereby forbid ad hoc tuning of model parameters, rendering the model more predictive.
In our calculation of $\mathrm{Re}(\epsilon'/\epsilon)$, we have considered all contributions
 including those from the left-right current operators, 
 unlike in our calculation of direct CP violation in $D\to PP$ decays.
In the plot, $\psi_d,\psi_s,\phi_c$ and $\alpha$ (which appears in Eq.~(\ref{wlwr})) 
 are randomly generated in the range $[0, 2\pi]$.
We observe that when the experimental value of $\mathrm{Re}(\epsilon'/\epsilon)$ is naturally accounted for,
 the CP asymmetry difference is about $10^{-6}$.
Conversely, to have $A_{\mathrm{CP}}^{D_s^+\to \pi^+\eta}-A_{\mathrm{CP}}^{D_s^+\to \pi^+\eta'}$
 as large as $10^{-4}$,
 one must fine-tune the new CP-violating phases to satisfy the $1\sigma$ range of $\mathrm{Re}(\epsilon'/\epsilon)$.
 
We comment in passing that for $W_R^+$ gauge boson in the $SU(2)_L\times SU(2)_R\times U(1)_{B-L}$ model,
 the constraint from indirect CP violation in kaons, Re$(\epsilon)$, is mild compared to that from direct CP violation $\mathrm{Re}(\epsilon'/\epsilon)$,
 because $W_R^+$ gauge boson exchange contributes to the latter at tree level while it contributes to the former only at loop levels. 
However, it should be noted that unless the scalar potential is fine-tuned,
 the contribution from the heavy neutral scalar exchange to Re$(\epsilon)$ is sizable, 
 which is investigated in detail in Refs.~\cite{Blanke:2011ry,Haba:2017jgf}.
\begin{table}[H]
\centering
\caption{Diagrammatic amplitudes~\cite{Bhattacharya:2009ps}, 
 process-dependent factors for CP asymmetry, and their numerical values
 for Cabibbo-favored charmed meson decays.
 The uncertainty comes from the $1\sigma$ errors
 of the diagrammatic amplitudes.
Here, we fix the $\eta-\eta^\prime$ mixing angle at $\arcsin(1/3)$.}
\label{Tab:1}
  \begin{tabular}{cccc} \hline\hline
  \rule[-9pt]{0pt}{26pt}
    $D\to f$ & ${\cal A}_f$&$F_{\mathrm{CP}}^{D\to f}$ &\# of $F_{\mathrm{CP}}^{D\to f}$ \\ \hline 
      \rule[-9pt]{0pt}{26pt}
    $D^+\to \bar{K^0}\pi^+$ & $T+C$
    &$0$ &$0$ \\ \hline 
          \rule[-9pt]{0pt}{26pt}
    $D^0\to K^-\pi^+$ & $T+E$ 
    &$\mathrm{Im}\left[\frac{C_\LL+A_\LL}{T_{\LL}+E_{\LL}}\right]$&$0.146\pm0.042$ \\ \hline 
          \rule[-9pt]{0pt}{26pt}
    $D^0\to \bar{K^0}\pi^0$ & $(C-E)/\sqrt{2}$
    &$\mathrm{Im}\left[\frac{T_\LL-A_\LL}{C_{\LL}-E_{\LL}}\right]$&$0.958\pm0.030$ \\ \hline 
          \rule[-9pt]{0pt}{26pt}
    $D^0\to \bar{K^0}\eta$ & $C/\sqrt{3}$ 
    &$\mathrm{Im}\left[\frac{
T_\LL}{C_{\LL}}\right]$& $0.595\pm0.015$ \\ \hline 
          \rule[-9pt]{0pt}{26pt}
    $D^0\to \bar{K^0}\eta^\prime$ & $-(C+3E)/\sqrt{6}$ 
    &$\mathrm{Im}\left[\frac{T_\LL+3A_\LL}
{C_{\LL}+3E_{\LL}}\right]$&$-0.479\pm0.076$ \\ \hline 
          \rule[-9pt]{0pt}{26pt}
    $D^+_s\to K^+\bar{K^0}$ & $C+A$ 
    &$\mathrm{Im}\left[\frac{T_\LL+E_\LL}{C_{\LL}+A_{\LL}}
\right]$&$-0.213\pm0.072$ \\ \hline 
          \rule[-9pt]{0pt}{26pt}
    $D^+_s\to \pi^+\eta$ & $(T-2A)/\sqrt{3}$ 
    &$\mathrm{Im}\left[\frac{C_\LL-2E_\LL}{T_{\LL}-2A_{\LL}}\right]$&$-1.367\pm0.074$ \\ \hline
         \rule[-9pt]{0pt}{26pt}
    $D^+_s\to \pi^+\eta^\prime$ & $2(T+A)/\sqrt{6}$ 
    &$\mathrm{Im}\left[\frac{
C_\LL+E_\LL}
{T_{\LL}+A_{\LL}}\right]$&$0.1726\pm0.039$\\ \hline\hline 
      \end{tabular}
\end{table}

\section{Summary}

We have studied the contribution of the right-right current operators in the 
 $SU(2)_L\times SU(2)_R\times U(1)_{B-L}$ model
 to direct CP violation in Cabibbo-favored charmed meson decays, for which the SM contribution is virtually absent.
Interestingly, this contribution is evaluable,
 because it stems from difference in QCD corrections to the left-left current operators induced by $W_L^+$ boson
 and the right-right ones induced by $W_R^+$ boson, 
 which is a short-distance effect $\sim(\alpha_s(M_{W_L}^2)/4\pi)\log(M_{W_R}^2/M_{W_L}^2)$.
Combining a short-distance calculation of this difference with
 the result of the diagrammatic approach to the Cabibbo-favored decay amplitudes,
 we numerically evaluate the CP asymmetry in
 $D^0\to K^-\pi^+$, $D^+_s\to \pi^+\eta$ and
$D^+_s\to \pi^+\eta^\prime$ decay rates.
We have found that the asymmetry in $D^+_s\to \pi^+\eta$ is specially sizable,
 and further suggested the measurement of the difference in the CP asymmetries in $D^+_s\to \pi^+\eta$ and
$D^+_s\to \pi^+\eta^\prime$ decays.
For $M_{W'}$ (almost equal to $M_{W_R}$) about 4~TeV and $g_R=2g_L$,
 one may hope to observe this CP asymmetry difference at Belle II with 50~ab$^{-1}$ of data or at LHCb with 50~fb$^{-1}$ of data.
Finally, we have presented a correlated prediction for the CP asymmetry difference in 
 $D^+_s\to \pi^+\eta$ and $D^+_s\to \pi^+\eta^\prime$ decays,
 and direct CP violation in $K\to \pi\pi$ decay Re$(\epsilon'/\epsilon)$, 
 under the assumption of `charge symmetry' in the $SU(2)_L\times SU(2)_R\times U(1)_{B-L}$ model.
We have observed that if the experimental data on Re$(\epsilon'/\epsilon)$ are naturally accounted for,
 the CP asymmetry difference in $D^+_s\to \pi^+\eta$ and $D^+_s\to \pi^+\eta^\prime$ decays is as small as $10^{-6}$,
 and that a fine-tuning of the new CP-violating phases
 is mandatory to anticipate the discovery of direct CP violation in Cabibbo-favored charmed meson decays.

\section*{Acknowledgement}
The authors would like to thank Monika~Blanke,
Jean-Marc~G\'erard,
Hsiang-nan~Li
and Farinaldo~S.~Queiroz
for useful comments.
This work is partially supported by Scientific Grants by the Ministry
of Education, Culture, Sports, Science and Technology of Japan
(Nos.~16H00871, 16H02189, 17K05415 and 18H04590).

\section*{Appendix A: NLO formulas}
Here, we summarize NLO QCD corrections
to the observables which are discussed in this paper.
At NLO, the ratio of the Wilson coefficients in Eq.~(\ref{wilsonratio})
is modified to
\bea
\left.\frac{C_\pm^{\mathrm{LL}}(\mu)}{C_\pm^{\mathrm{RR}}(\mu)}\right|_{\mathrm{NLO}}
&=&
U^{\mathrm{NLO}}_\pm
\left(\frac{g_R}{g_L}\frac{M_W}{M_{W^\prime}}\right)^2
\frac{V_{cs}^{\mathrm{R}*}V_{ud}^{\mathrm{R}}}{V_{cs}^{\mathrm{L}*}V_{ud}^{\mathrm{L}}},\label{Eq:ratio}\qquad\\
U^{\mathrm{NLO}}_\pm&=&(\eta_{\mu_{W^\prime}}^{\mu_{W}})^{-\frac{\lambda_{0\pm}}{2\beta_0}}
\left[1-\frac{\alpha_s(\mu_W)}{4\pi}
\left(\frac{\beta_1\lambda_{0\pm}}{2\beta_0^2}-\frac{\lambda_{1\pm}}{2\beta_0}+\frac{\lambda_{0\pm}}{2}\log\frac{M_W^2}{\mu_W^2}-B_\pm\right)\right]
\nonumber\\
&\times&\left[1+\frac{\alpha_s(\mu_{W'})}{4\pi}
\left(\frac{\beta_1\lambda_{0\pm}}{2\beta_0^2}-\frac{\lambda_{1\pm}}{2\beta_0}+\frac{\lambda_{0\pm}}{2}\log\frac{M_{W'}^2}{\mu_{W'}^2}-B_\pm\right)\right],
\label{Eq:asymme}
\eea
where $\beta_1$ is the six-flavor NLO QCD $\beta$ function coefficient,
$\lambda_{1\pm}$ are the NLO $\gamma$ function coefficients 
for $C^{\rm LL}_\pm$ and $C^{\rm RR}_\pm$,
and $B_\pm$ are constants
(see, \textit{e.g.}, Ref.~\cite{Buras:1998raa}).
Note that each of $\lambda_{1\pm}$ and $B_\pm$ is renormalization-scheme-dependent, 
but their scheme dependences cancel.
Thus at NLO,
the CP asymmetry in Eq.~(\ref{Eq:ASS}) is
\bea
A_{\mathrm{CP}}^{D\to f}|_{\mathrm{NLO}}&=&F_{\mathrm{CP}}^{D\to f}
\left[U_+^{\mathrm{NLO}}
-U_-^{\mathrm{NLO}}
\right]\left(\frac{g_R}{g_L}\frac{M_W}{M_{W'}}\right)^2
\mathrm{Im}\left(\frac{V_{cs}^{\mathrm{R}*}V_{ud}^{\mathrm{R}}}
{V_{cs}^{\mathrm{L}*}V_{ud}^{\mathrm{L}}}\label{NLO}
\right).
\eea
\\

\section*{Appendix B: Effect of $SU(3)$ flavor symmetry breaking}

We study the effect of $SU(3)$ flavor symmetry breaking on our prediction,
which is not discussed in the main text.
Our prediction of CP asymmetries
 depends crucially on V-spin symmetry 
 (symmetry of $u$ and $s$, which is part of $SU(3)$ flavor symmetry),
 since our prediction is derived from Eqs.~(\ref{ce}--\ref{da}),
 which are obtained by assuming V-spin.
In particular, the isospin symmetry cannot lead to the above results.
The effect of V-spin breaking on Eqs.~(\ref{ce}--\ref{da}) 
 is estimated to be simply $f_K/f_\pi-1\simeq0.2$.
This is in contrast to singly-Cabibbo-suppressed decays,
 where $SU(3)$ breaking gives rise to
 corrections of order $(f_K/f_\pi)^2-1\simeq0.4$ in factorized tree amplitudes,
 and also enhances penguin amplitudes (suppressed by $V_{cb}V_{ub}^*$ in the $SU(3)$ limit)
 leading to a further splitting of $c\to d\bar{d}u$-induced amplitudes
 and $c\to s\bar{s}u$-induced amplitudes~\cite{Bhattacharya:2012kq};
 all these effects are absent in the Cabibbo-favored decays.

Let us see how corrections of order $f_K/f_\pi-1\simeq0.2$ to Eqs.~(\ref{ce}--\ref{da})
 affect our prediction of CP asymmetries.
First we concentrate on $T$(tree) and $C$(color-suppressed tree) diagrams.
When Eqs.~(\ref{ce},\ref{de}) are not valid, $T_{\LL}$ and $C_{\LL}$ are written as
\bea
T_{\LL}&=&C_{+}^{\LL}\frac{\braket{Q_1}_{\mathrm{CE}}+\braket{Q_2}_{\mathrm{DE}}}{2}
+C_{-}^{\LL}\frac{\braket{Q_1}_{\mathrm{CE}}-\braket{Q_2}_{\mathrm{DE}}}{2},
\label{newt}\\
C_{\LL}&=&C_{+}^{\LL}\frac{\braket{Q_2}_{\mathrm{CE}}+\braket{Q_1}_{\mathrm{DE}}}{2}
-C_{-}^{\LL}\frac{\braket{Q_2}_{\mathrm{CE}}-\braket{Q_1}_{\mathrm{DE}}}{2}.
\label{newc}
\eea
The first and second terms on the right-hand side of Eqs.~(\ref{newt}, \ref{newc})
 are individually renormalization-scale-and-scheme independent.
Therefore, we can parametrize the V-spin breaking effects in terms of
 renormalization-scale-and-scheme independent parameters $\epsilon_{\rm{E}+}$ and 
 $\epsilon_{\rm{E}-}$ as
\bea
C_{+}^{\LL}\frac{\braket{Q_1}_{\mathrm{CE}}+\braket{Q_2}_{\mathrm{DE}}}{2}
&=&
C_{+}^{\LL}\frac{\braket{Q_2}_{\mathrm{CE}}+\braket{Q_1}_{\mathrm{DE}}}{2}(1+\epsilon_{\rm{E}+}),
\\
C_{-}^{\LL}\frac{\braket{Q_1}_{\mathrm{CE}}-\braket{Q_2}_{\mathrm{DE}}}{2}
&=&
C_{-}^{\LL}\frac{\braket{Q_2}_{\mathrm{CE}}-\braket{Q_1}_{\mathrm{DE}}}{2}(1+\epsilon_{\rm{E}-}),
\eea
 where we estimate the V-spin breaking parameters as
$\vert\epsilon_{\rm{E}+}\vert\sim\vert\epsilon_{\rm{E}-}\vert\sim f_K/f_\pi-1\simeq0.2$.
In the leading order of $\epsilon_{\rm{E}+},\epsilon_{\rm{E}-}$, we find
\bea
T_{\LL}+(1+\epsilon_{\rm{E}-})C_{\LL}=C_{+}^{\LL}
(\braket{Q_2}_{\rm{CE}}+\braket{Q_1}_{\rm{DE}})(1+\epsilon_{\rm{E}-}/2+\epsilon_{\rm{E}+}/2),
\\
T_{\LL}-(1+\epsilon_{\rm{E}+})C_{\LL}=C_{-}^{\LL}
(\braket{Q_2}_{\rm{CE}}-\braket{Q_1}_{\rm{DE}})(1+\epsilon_{\rm{E}-}/2+\epsilon_{\rm{E}+}/2).
\eea
Consequently, $T_{\RR}$ and $C_{\RR}$ can be expressed in terms of $T_{\LL}$, $C_{\LL}$ and the V-spin breaking parameters as
\bea
T_{\RR}&=&-\frac{C_{+}^{\RR}}{C_{+}^{\LL}}\frac{T_{\LL}(1+\epsilon_{\rm{E}+}/2-\epsilon_{\rm{E}-}/2)
+C_{\LL}(1+\epsilon_{\rm{E}+}/2+\epsilon_{\rm{E}-}/2)}{2}
\nn\\
&-&\frac{C_{-}^{\RR}}{C_{-}^{\LL}}\frac{T_{\LL}(1-\epsilon_{\rm{E}+}/2+\epsilon_{\rm{E}-}/2)
-C_{\LL}(1+\epsilon_{\rm{E}+}/2+\epsilon_{\rm{E}-}/2)}{2},
\label{trr2}\\
C_{\RR}&=&-\frac{C_{+}^{\RR}}{C_{+}^{\LL}}\frac{T_{\LL}(1-\epsilon_{\rm{E}+}/2-\epsilon_{\rm{E}-}/2)
+C_{\LL}(1-\epsilon_{\rm{E}+}/2+\epsilon_{\rm{E}-}/2)}{2}
\nn\\
&+&\frac{C_{-}^{\RR}}{C_{-}^{\LL}}\frac{T_{\LL}(1-\epsilon_{\rm{E}+}/2-\epsilon_{\rm{E}-}/2)
-C_{\LL}(1+\epsilon_{\rm{E}+}/2-\epsilon_{\rm{E}-}/2)}{2}.
\label{crr2}
\eea
We obtain analogous expressions for $A_{\RR}$ and $E_{\RR}$, with $\epsilon_{\rm{E}+},\epsilon_{\rm{E}-}$
 replaced with different V-spin breaking parameters $\epsilon_{\rm{A}+},\epsilon_{\rm{A}-}$.
The above V-spin breaking corrections solely affect the factor $F_{\mathrm{CP}}^{D\to f}$ in the formula for CP asymmetry Eq.~(\ref{Eq:ASS}).
For the phenomenologically interesting modes $D_s^+\to\pi^+\eta$, $D_s^+\to\pi^+\eta'$ and $D^0\to K^-\pi^+$,
 this factor is altered from Table~\ref{Tab:1} to
\bea
F_{\rm CP}^{D_s^+\to\pi^+\eta}&=&\rm{Im}\left[\frac{C_{\LL}-2E_{\LL}}{T_{\LL}-2A_{\LL}}\right]
+{\rm Im}\left[\frac{C_{\LL}(\epsilon_{\rm{E}+}/2+\epsilon_{\rm{E}-}/2)-2E_{\LL}(\epsilon_{\rm{A}+}/2+\epsilon_{\rm{A}-}/2)}
{T_{\LL}-2A_{\LL}}\right]
\nn\\
&+&{\rm Im}\left[\frac{T_{\LL}(\epsilon_{\rm{E}+}/2-\epsilon_{\rm{E}-}/2)-2A_{\LL}(\epsilon_{\rm{A}+}/2-\epsilon_{\rm{A}-}/2)}
{T_{\LL}-2A_{\LL}}\right],
\label{newf1}\\
F_{\rm CP}^{D_s^+\to\pi^+\eta'}&=&\rm{Im}\left[\frac{C_{\LL}+E_{\LL}}{T_{\LL}+A_{\LL}}\right]
+{\rm Im}\left[\frac{C_{\LL}(\epsilon_{\rm{E}+}/2+\epsilon_{\rm{E}-}/2)+E_{\LL}(\epsilon_{\rm{A}+}/2+\epsilon_{\rm{A}-}/2)}
{T_{\LL}+A_{\LL}}\right]
\nn\\
&+&{\rm Im}\left[\frac{T_{\LL}(\epsilon_{\rm{E}+}/2-\epsilon_{\rm{E}-}/2)+A_{\LL}(\epsilon_{\rm{A}+}/2-\epsilon_{\rm{A}-}/2)}
{T_{\LL}+A_{\LL}}\right],
\label{newf2}\\
F_{\rm CP}^{D^0\to K^-\pi^+}&=&\rm{Im}\left[\frac{C_{\LL}+A_{\LL}}{T_{\LL}+E_{\LL}}\right]
+{\rm Im}\left[\frac{C_{\LL}(\epsilon_{\rm{E}+}/2+\epsilon_{\rm{E}-}/2)+A_{\LL}(-\epsilon_{\rm{A}+}/2-\epsilon_{\rm{A}-}/2)}
{T_{\LL}+E_{\LL}}\right]
\nn\\
&+&{\rm Im}\left[\frac{T_{\LL}(\epsilon_{\rm{E}+}/2-\epsilon_{\rm{E}-}/2)+E_{\LL}(-\epsilon_{\rm{A}+}/2+\epsilon_{\rm{A}-}/2)}
{T_{\LL}+E_{\LL}}\right].
\label{newf3}
\eea
Depending on the phases of V-spin breaking parameters $\epsilon_{\rm{E}+},\epsilon_{\rm{E}-},\epsilon_{\rm{A}+},\epsilon_{\rm{A}-}$,
 the second and third terms of Eqs.~(\ref{newf1}--\ref{newf3}) can be enhanced far beyond $f_K/f_\pi-1\simeq0.2$.
However, as we will show below, the most promising observable,
 $A_{\mathrm{CP}}^{D_s^+\to \pi^+\eta}-A_{\mathrm{CP}}^{D_s^+\to \pi^+\eta'}$, is not much affected by the V-spin breaking.
To see this, note that this observable is proportional to $F_{\mathrm{CP}}^{D_s^+\to \pi^+\eta}-F_{\mathrm{CP}}^{D_s^+\to \pi^+\eta'}$.
Since $\vert A_{\LL}\vert$ is small, it can be approximated as
\bea
F_{\mathrm{CP}}^{D_s^+\to \pi^+\eta}-F_{\mathrm{CP}}^{D_s^+\to \pi^+\eta'}
&=&\rm{Im}\left[\frac{C_{\LL}-2E_{\LL}}{T_{\LL}-2A_{\LL}}\right]-\rm{Im}\left[\frac{C_{\LL}+E_{\LL}}{T_{\LL}+A_{\LL}}\right]
\nn\\
&&-3\,{\rm Im}\left[\frac{E_{\LL}(\epsilon_{\rm{A}+}/2+\epsilon_{\rm{A}-}/2)}{T_{\LL}}\right]
-3\,{\rm Im}\left[\frac{A_{\LL}(\epsilon_{\rm{A}+}/2-\epsilon_{\rm{A}-}/2)}{T_{\LL}}\right]
\\
&=&-1.54
-3\,{\rm Im}\left[\frac{E_{\LL}(\epsilon_{\rm{A}+}/2+\epsilon_{\rm{A}-}/2)}{T_{\LL}}\right]
-3\,{\rm Im}\left[\frac{A_{\LL}(\epsilon_{\rm{A}+}/2-\epsilon_{\rm{A}-}/2)}{T_{\LL}}\right],
\nn\\\label{newfdifference}
\eea
 where the first term $-1.54$ is the prediction in the V-spin limit, while the second and third terms represent V-spin breaking effects.
The second term is at most $\pm3\vert E_{\LL}/T_{\LL}\vert(f_K/f_\pi-1)\simeq\pm0.32$ 
 and the third term is at most $\pm3\vert A_{\LL}/T_{\LL}\vert(f_K/f_\pi-1)\simeq\pm0.07$.
Thus, we conclude that the V-spin breaking corrections do not significantly change our prediction of
$F_{\mathrm{CP}}^{D_s^+\to \pi^+\eta}-F_{\mathrm{CP}}^{D_s^+\to \pi^+\eta'}$ and hence of
$A_{\mathrm{CP}}^{D_s^+\to \pi^+\eta}-A_{\mathrm{CP}}^{D_s^+\to \pi^+\eta'}$.

We note in passing that the fitted values of $T_{\LL},C_{\LL},E_{\LL},A_{\LL}$ in Ref.~\cite{Bhattacharya:2009ps}, 
 which we have adopted throughout the paper, 
 are themselves obtained under the assumption of $SU(3)$ flavor symmetry,
 but we expect that the $SU(3)$ breaking effects are properly
 reflected in the errors of the fitted values.



\begin{thebibliography}{99}
\bibitem{Delepine:2012xw} 
  D.~Delepine, G.~Faisel and C.~A.~Ramirez,
  ``Observation of CP violation in $D^0\to K^-\pi^+$ as a smoking gun for new physics,''
  Phys.\ Rev.\ D {\bf 87}, no. 7, 075017 (2013)
  [arXiv:1212.6281 [hep-ph]].

\bibitem{Golden:1989qx} 
  M.~Golden and B.~Grinstein,
  ``Enhanced CP Violations in Hadronic Charm Decays,''
  Phys.\ Lett.\ B {\bf 222}, 501 (1989).

\bibitem{Bigi:1994aw} 
  I.~I.~Y.~Bigi and H.~Yamamoto,
  ``Interference between Cabibbo allowed and doubly forbidden transitions in $D \to K_{S, L}+\pi's$ decays,''
  Phys.\ Lett.\ B {\bf 349}, 363 (1995)
  [hep-ph/9502238].
  
\bibitem{Xing:1995jg} 
  Z.~Z.~Xing,
  ``Effect of $K^0-\bar{K^0}$ mixing on CP asymmetries in weak decays of $D$ and $B$ mesons,''
  Phys.\ Lett.\ B {\bf 353}, 313 (1995)
  Erratum: [Phys.\ Lett.\ B {\bf 363}, 266 (1995)]
  [hep-ph/9505272].
 
    
\bibitem{Haba:2018byj} 
  N.~Haba, H.~Umeeda and T.~Yamada,
  ``$\epsilon'/\epsilon$ Anomaly and Neutron EDM in $SU(2)_L\times SU(2)_R\times U(1)_{B-L}$ model with Charge Symmetry,''
  JHEP {\bf 1805}, 052 (2018)
  [arXiv:1802.09903 [hep-ph]].

 
\bibitem{Buras:1998ra} 
  A.~J.~Buras and L.~Silvestrini,
  ``Nonleptonic two-body B decays beyond factorization,''
  Nucl.\ Phys.\ B {\bf 569}, 3 (2000)
  [hep-ph/9812392].
 
 
\bibitem{Zeppenfeld:1980ex} 
  D.~Zeppenfeld,
  ``SU(3) Relations for $B$ Meson Decays,''
  Z.\ Phys.\ C {\bf 8}, 77 (1981).
  
\bibitem{Chau:1982da} 
  L.~L.~Chau,
  ``Quark Mixing in Weak Interactions,''
  Phys.\ Rept.\  {\bf 95}, 1 (1983).
  
\bibitem{Gronau:1994bn} 
  M.~Gronau, J.~L.~Rosner and D.~London,
  ``Weak coupling phase from decays of charged $B$ mesons to pi K and pi pi,''
  Phys.\ Rev.\ Lett.\  {\bf 73}, 21 (1994)
  [hep-ph/9404282].
  
\bibitem{Hernandez:1994rh} 
  O.~F.~Hernandez, D.~London, M.~Gronau and J.~L.~Rosner,
  ``Measuring strong and weak phases in time independent B decays,''
  Phys.\ Lett.\ B {\bf 333}, 500 (1994)
  [hep-ph/9404281].
  
\bibitem{Gronau:1994rj} 
  M.~Gronau, O.~F.~Hernandez, D.~London and J.~L.~Rosner,
  ``Decays of $B$ mesons to two light pseudoscalars,''
  Phys.\ Rev.\ D {\bf 50}, 4529 (1994)
  [hep-ph/9404283].
 
\bibitem{Bhattacharya:2009ps} 
  B.~Bhattacharya and J.~L.~Rosner,
  ``Charmed meson decays to two pseudoscalars,''
  Phys.\ Rev.\ D {\bf 81}, 014026 (2010)
  [arXiv:0911.2812 [hep-ph]].

\bibitem{Rosner:1999xd} 
  J.~L.~Rosner,
  ``Final state phases in charmed meson two-body nonleptonic decays,''
  Phys.\ Rev.\ D {\bf 60}, 114026 (1999)
  [hep-ph/9905366].


\bibitem{Bhattacharya:2008ss} 
  B.~Bhattacharya and J.~L.~Rosner,
  ``Flavor symmetry and decays of charmed mesons to pairs of light pseudoscalars,''
  Phys.\ Rev.\ D {\bf 77}, 114020 (2008)
  [arXiv:0803.2385 [hep-ph]].
 
 
\bibitem{Chau:1986du} 
  L.~L.~Chau and H.~Y.~Cheng,
  ``Quark Diagram Analysis of Two-body Charm Decays,''
  Phys.\ Rev.\ Lett.\  {\bf 56}, 1655 (1986).
  
\bibitem{Chau:1987tk} 
  L.~L.~Chau and H.~Y.~Cheng,
  ``Analysis of Exclusive Two-Body Decays of Charm Mesons Using the Quark Diagram Scheme,''
  Phys.\ Rev.\ D {\bf 36}, 137 (1987)
  Addendum: [Phys.\ Rev.\ D {\bf 39}, 2788 (1989)].
  
\bibitem{Chau:1989tk} 
  L.~L.~Chau and H.~Y.~Cheng,
  ``Analysis of the Recent Data of Exclusive Two-body Charm Decays,''
  Phys.\ Lett.\ B {\bf 222}, 285 (1989).
 
\bibitem{Cheng:2010ry} 
  H.~Y.~Cheng and C.~W.~Chiang,
  ``Two-body hadronic charmed meson decays,''
  Phys.\ Rev.\ D {\bf 81}, 074021 (2010)
  [arXiv:1001.0987 [hep-ph]].
  
   
\bibitem{Maiezza:2010ic} 
  A.~Maiezza, M.~Nemevsek, F.~Nesti and G.~Senjanovic,
  ``Left-Right Symmetry at LHC,''
  Phys.\ Rev.\ D {\bf 82}, 055022 (2010)
  [arXiv:1005.5160 [hep-ph]].
 
 
  \bibitem{Batley:2002gn} 
  J.~R.~Batley {\it et al.} [NA48 Collaboration],
  ``A Precision measurement of direct CP violation in the decay of neutral kaons into two pions,''
  Phys.\ Lett.\ B {\bf 544}, 97 (2002)
  [hep-ex/0208009].
\bibitem{AlaviHarati:2002ye} 
  A.~Alavi-Harati {\it et al.} [KTeV Collaboration],
  ``Measurements of direct CP violation, CPT symmetry, and other parameters in the neutral kaon system,''
  Phys.\ Rev.\ D {\bf 67}, 012005 (2003)
  Erratum: [Phys.\ Rev.\ D {\bf 70}, 079904 (2004)]
  [hep-ex/0208007].
\bibitem{Abouzaid:2010ny} 
  E.~Abouzaid {\it et al.} [KTeV Collaboration],
  ``Precise Measurements of Direct CP Violation, CPT Symmetry, and Other Parameters in the Neutral Kaon System,''
  Phys.\ Rev.\ D {\bf 83}, 092001 (2011)
  [arXiv:1011.0127 [hep-ex]].
  
  
 
\bibitem{Buras:2015xba} 
  A.~J.~Buras and J.~M.~G\'erard,
  ``Upper bounds on ε′/ε parameters B$_{6}^{(1/2)}$ and B$_{8}^{(3/2)}$ from large N QCD and other news,''
  JHEP {\bf 1512}, 008 (2015)
  [arXiv:1507.06326 [hep-ph]].
\bibitem{Buras:2016fys} 
  A.~J.~Buras and J.~M.~G\'erard,
  ``Final state interactions in $K\rightarrow \pi \pi $ decays: $\Delta I=1/2$ rule vs. $\varepsilon '/\varepsilon $,''
  Eur.\ Phys.\ J.\ C {\bf 77}, no. 1, 10 (2017)
  [arXiv:1603.05686 [hep-ph]].
 

  \bibitem{Blum:2011ng} 
  T.~Blum {\it et al.},
  ``The $K\to(\pi\pi)_{I=2}$ Decay Amplitude from Lattice QCD,''
  Phys.\ Rev.\ Lett.\  {\bf 108}, 141601 (2012)
  [arXiv:1111.1699 [hep-lat]];
  ``Lattice determination of the $K \to (\pi\pi)_{I=2}$ Decay Amplitude $A_2$,''
  Phys.\ Rev.\ D {\bf 86}, 074513 (2012)
  [arXiv:1206.5142 [hep-lat]].
\bibitem{Blum:2015ywa} 
  T.~Blum {\it et al.},
  ``$K \rightarrow \pi\pi$ $\Delta I=3/2$ decay amplitude in the continuum limit,''
  Phys.\ Rev.\ D {\bf 91}, no. 7, 074502 (2015)
  [arXiv:1502.00263 [hep-lat]].
\bibitem{Bai:2015nea} 
  Z.~Bai {\it et al.} [RBC and UKQCD Collaborations],
  ``Standard Model Prediction for Direct CP Violation in K→ππ Decay,''
  Phys.\ Rev.\ Lett.\  {\bf 115}, no. 21, 212001 (2015)
  [arXiv:1505.07863 [hep-lat]].
  \bibitem{Buras:2015yba} 
  A.~J.~Buras, M.~Gorbahn, S.~J\"ager and M.~Jamin,
  ``Improved anatomy of $\epsilon^\prime/\epsilon$ in the Standard Model,''
  JHEP {\bf 1511}, 202 (2015)
  [arXiv:1507.06345 [hep-ph]].
\bibitem{Kitahara:2016nld} 
  T.~Kitahara, U.~Nierste and P.~Tremper,
  ``Singularity-free next-to-leading order $\Delta$S = 1 renormalization group evolution and $\epsilon_K'/\epsilon_K$ in the Standard Model and beyond,''
  JHEP {\bf 1612}, 078 (2016)
  [arXiv:1607.06727 [hep-ph]].

  
  
\bibitem{Cirigliano:2016yhc} 
  V.~Cirigliano, W.~Dekens, J.~de Vries and E.~Mereghetti,
  ``An $\epsilon^\prime$ improvement from right-handed currents,''
  Phys.\ Lett.\ B {\bf 767}, 1 (2017)
  [arXiv:1612.03914 [hep-ph]].
\bibitem{Blanke:2015wba} 
  M.~Blanke, A.~J.~Buras and S.~Recksiegel,
  ``Quark flavour observables in the Littlest Higgs model with T-parity after LHC Run 1,''
  Eur.\ Phys.\ J.\ C {\bf 76}, no. 4, 182 (2016)
  [arXiv:1507.06316 [hep-ph]].
\bibitem{Tanimoto:2016yfy} 
  M.~Tanimoto and K.~Yamamoto,
  ``Probing SUSY with 10 TeV stop mass in rare decays and CP violation of kaon,''
  PTEP {\bf 2016}, no. 12, 123B02 (2016)
  [arXiv:1603.07960 [hep-ph]].
\bibitem{Kitahara:2016otd} 
  T.~Kitahara, U.~Nierste and P.~Tremper,
  ``Supersymmetric Explanation of CP Violation in $K\to \pi\pi$ Decays,''
  Phys.\ Rev.\ Lett.\  {\bf 117}, no. 9, 091802 (2016)
  [arXiv:1604.07400 [hep-ph]].
\bibitem{Endo:2016aws} 
  M.~Endo, S.~Mishima, D.~Ueda and K.~Yamamoto,
  ``Chargino contributions in light of recent $\epsilon'/\epsilon$,''
  Phys.\ Lett.\ B {\bf 762}, 493 (2016)
  [arXiv:1608.01444 [hep-ph]].
\bibitem{Buras:2015jaq} 
  A.~J.~Buras,
  ``New physics patterns in $\varepsilon^\prime/\varepsilon$ and $\varepsilon_K$ with implications for rare kaon decays and $\Delta M_K$,''
  JHEP {\bf 1604}, 071 (2016)
  [arXiv:1601.00005 [hep-ph]].
\bibitem{Endo:2016tnu} 
  M.~Endo, T.~Kitahara, S.~Mishima and K.~Yamamoto,
  ``Revisiting Kaon Physics in General $Z$ Scenario,''
  Phys.\ Lett.\ B {\bf 771}, 37 (2017)
  [arXiv:1612.08839 [hep-ph]].
\bibitem{Bobeth:2016llm} 
  C.~Bobeth, A.~J.~Buras, A.~Celis and M.~Jung,
  ``Patterns of Flavour Violation in Models with Vector-Like Quarks,''
  JHEP {\bf 1704}, 079 (2017)
  [arXiv:1609.04783 [hep-ph]].
\bibitem{Buras:2015kwd} 
  A.~J.~Buras and F.~De Fazio,
  ``$\varepsilon'/\varepsilon$ in 331 Models,''
  JHEP {\bf 1603}, 010 (2016)
  [arXiv:1512.02869 [hep-ph]];
  ``331 Models Facing the Tensions in $\Delta F=2$ Processes with the Impact on $\varepsilon^\prime/\varepsilon$, $B_s\to\mu^+\mu^-$ and $B\to K^*\mu^+\mu^-$,''
  JHEP {\bf 1608}, 115 (2016)
  [arXiv:1604.02344 [hep-ph]].
\bibitem{Chen:2018ytc} 
  C.~H.~Chen and T.~Nomura,
  ``$\epsilon'_K/\epsilon_K$ and $K \to \pi \nu \bar\nu$ in a two-Higgs doublet model,''
  arXiv:1804.06017 [hep-ph].
\bibitem{Chen:2018vog} 
  C.~H.~Chen and T.~Nomura,
  ``$\epsilon'/\epsilon$ from charged-Higgs-induced gluonic dipole operators,''
  arXiv:1805.07522 [hep-ph].
\bibitem{Matsuzaki:2018jui} 
  S.~Matsuzaki, K.~Nishiwaki and K.~Yamamoto,
  ``Simultaneous interpretation of $K$ and $B$ anomalies in terms of chiral-flavorful vectors,''
  arXiv:1806.02312 [hep-ph].

\bibitem{Won:2011ku} 
  E.~Won {\it et al.} [Belle Collaboration],
  ``Observation of $D^+ \rightarrow K^{+} \eta^{(\prime)}$ and Search for CP Violation in $D^+ \rightarrow \pi^+ \eta^{(\prime)}$ Decays,''
  Phys.\ Rev.\ Lett.\  {\bf 107}, 221801 (2011)
  [arXiv:1107.0553 [hep-ex]].

  
\bibitem{Patrignani:2016xqp} 
  C.~Patrignani {\it et al.} [Particle Data Group],
  ``Review of Particle Physics,''
  Chin.\ Phys.\ C {\bf 40}, no. 10, 100001 (2016).


\bibitem{Aaij:2017eux} 
  R.~Aaij {\it et al.} [LHCb Collaboration],
  ``Measurement of $CP$ asymmetries in $D^{\pm}\rightarrow \eta^{\prime} \pi^{\pm}$ and $D_s^{\pm}\rightarrow \eta^{\prime} \pi^{\pm}$ decays,''
  Phys.\ Lett.\ B {\bf 771}, 21 (2017)
  [arXiv:1701.01871 [hep-ex]].

\bibitem{Bediaga:2012py}
  R.~Aaij {\it et al.} [LHCb Collaboration],
  ``Implications of LHCb measurements and future prospects,''
  Eur.\ Phys.\ J.\ C {\bf 73} (2013) no.4,  2373
  [arXiv:1208.3355 [hep-ex]].

\bibitem{Li:2012cfa} 
  H.~n.~Li, C.~D.~Lu and F.~S.~Yu,
  ``Branching ratios and direct CP asymmetries in $D\to PP$ decays,''
  Phys.\ Rev.\ D {\bf 86}, 036012 (2012)
  [arXiv:1203.3120 [hep-ph]].


\bibitem{Patra:2015bga} 
  S.~Patra, F.~S.~Queiroz and W.~Rodejohann,
  ``Stringent Dilepton Bounds on Left-Right Models using LHC data,''
  Phys.\ Lett.\ B {\bf 752}, 186 (2016)
  [arXiv:1506.03456 [hep-ph]].
\bibitem{Lindner:2016lpp} 
  M.~Lindner, F.~S.~Queiroz and W.~Rodejohann,
  ``Dilepton bounds on left-right symmetry at the LHC run II and neutrinoless double beta decay,''
  Phys.\ Lett.\ B {\bf 762}, 190 (2016)
  [arXiv:1604.07419 [hep-ph]].
  
\bibitem{Blanke:2011ry} 
  M.~Blanke, A.~J.~Buras, K.~Gemmler and T.~Heidsieck,
  ``$\Delta F = 2$ observables and $B \to X_q\gamma$  decays in the Left-Right Model: Higgs particles striking back,''
  JHEP {\bf 1203}, 024 (2012)
  [arXiv:1111.5014 [hep-ph]].
  
\bibitem{Haba:2017jgf} 
  N.~Haba, H.~Umeeda and T.~Yamada,
  ``Semialigned two Higgs doublet model,''
  Phys.\ Rev.\ D {\bf 97}, no. 3, 035004 (2018)
  [arXiv:1711.06499 [hep-ph]].

\bibitem{Buras:1998raa} 
A.~J.~Buras,
  ``Weak Hamiltonian, CP violation and rare decays,''
  hep-ph/9806471.
  
\bibitem{Bhattacharya:2012kq} 
  B.~Bhattacharya, M.~Gronau and J.~L.~Rosner,
  ``Direct CP Violation in D Decays in view of LHCb and CDF Results,''
  arXiv:1207.0761 [hep-ph].

 \end{thebibliography}
\end{document}